\RequirePackage{silence}
\documentclass[fleqn,usenatbib,useAMS]{mnras} 
\usepackage{graphicx}
\usepackage{amsmath}

\usepackage{amsfonts}
\usepackage{float}
\usepackage{bm}
\usepackage[perpage,symbol*]{footmisc}
\usepackage{ae,aecompl}
\usepackage{array}
\usepackage{soul}
\usepackage{mathtools}
\usepackage{multirow}

\newcommand{\appropto}{\mathrel{\vcenter{
		\offinterlineskip\halign{\hfil$##$\cr 
			\propto\cr\noalign{\kern2pt}\sim\cr\noalign{\kern-2pt}}}}}

\title[Testing gravity on interstellar precursor missions]{Testing gravity with interstellar precursor missions} 

\author[Indranil Banik \& Pavel Kroupa]{Indranil Banik$^{1}$\thanks{Email:
\href{mailto:ibanik@astro.uni-bonn.de}{ibanik@astro.uni-bonn.de} (Indranil Banik)\newline $~~~~~~~~~~~~~~~~~$ \href{mailto:pavel@astro.uni-bonn.de}{pavel@astro.uni-bonn.de} (Pavel Kroupa)} and Pavel Kroupa$^{1,2}$\\
$^{1}$Helmholtz-Institut f\"ur Strahlen und Kernphysik (HISKP), University of Bonn, Nussallee 14$-$16, D-53115 Bonn, Germany \\
$^{2}$Charles University, Faculty of Mathematics and Physics, Astronomical Institute, V Hole\v{s}ovi\v{c}k\'ach 2, CZ-18000 Praha 8, Czech Republic}

\pubyear{2019}
\begin{document}
\label{firstpage}
\pagerange{\pageref{firstpage}--\pageref{lastpage}}

\maketitle

\begin{abstract}

We consider how the trajectory of an interstellar precursor mission would be affected by the gravity of the Sun in Newtonian and Milgromian dynamics (MOND). The Solar gravity is ${\approx 50\%}$ stronger in MOND beyond a distance of ${\approx 7000}$ astronomical units, the MOND radius of the Sun. A spacecraft travelling at $0.01 \, c$ reaches this distance after 11.1 years. We show that the extra gravity in MOND causes an anomalous deceleration that reduces its radial velocity by ${\approx 3}$ cm/s and the two-way light travel time from the inner Solar System by ${\approx 0.1}$ seconds after 20 years. A distinctive signature of MOND is that the gravity from the Sun is not directly towards it. This is due to the non-linear nature of MOND and the external gravitational field from the rest of the Galaxy, which we self-consistently include in our calculations. As a result, the sky position of the spacecraft would deviate by up to 0.2 mas over 20 years. This deviation is always in the plane containing the spacecraft trajectory and the direction towards the Galactic Centre. By launching spacecraft in different directions, it is possible to test the characteristic pattern of angular deviations expected in MOND. This would minimize the chance that any detected anomalies are caused by other processes like drag from the interstellar medium. Such confounding factors could also be mitigated using an onboard accelerometer to measure non-gravitational forces. We briefly discuss how the gravity theories could be conclusively distinguished using a Cavendish-style active gravitational experiment beyond the Sun's MOND radius.

\end{abstract}


\section{Introduction}
\label{Introduction}

The Breakthrough Starshot initiative has prompted a closer look at the challenges involved in reaching another star \citep{Merali_2016}. Along the way, much exciting science will become possible, for instance dramatically improved trigonometric parallaxes to stars across the Local Group \citep{Jaffe_1979, Etchegaray_1987}. Here, we focus on the fact that several thousand astronomical units (several kAU) from the Sun, its gravitational field is weaker than the gravity typically experienced by stars and gas in the outskirts of galaxies. Such matter behaves in a very unusual way: galaxy rotation curves are asymptotically flat instead of following the expected Keplerian decline beyond the extent of their luminous matter \citep[e.g.][]{Babcock_1939, Rubin_Ford_1970, Rogstad_1972, Roberts_1975}.

These acceleration discrepancies are conventionally attributed to halos of cold dark matter surrounding each galaxy \citep{Ostriker_Peebles_1973}. However, this has led to several problems \citep{Kroupa_2012, Kroupa_2015}. In the Local Group, the most serious are the existence of satellite galaxy planes \citep{Pawlowski_2018} and high-velocity dwarfs \citep{Banik_2017_anisotropy}. Both problems are rather immune to baryonic physics because they involve length scales of several hundred kpc and relate to the motion of galaxies as a whole. This was recently demonstrated for the satellite plane problem by approximately including the main way in which baryonic physics affects it \citep{Pawlowski_2019}. A highly anisotropic satellite system also exists around Centaurus A \citep{Muller_2018}.

These problems might indicate that galaxies are actually governed by Milgromian dynamics \citep[MOND,][]{Milgrom_1983}. In MOND, the gravitational field strength $g$ at distance $r$ from an isolated point mass $M$ transitions from the Newtonian ${GM/r^2}$ law at short range to
\begin{eqnarray}
	g ~=~ \frac{\sqrt{GMa_{_0}}}{r} ~~~\text{for } ~ r \gg \overbrace{\sqrt{\frac{GM}{a_{_0}}}}^{r_{_M}} \, .
	\label{Deep_MOND_limit}
\end{eqnarray}
MOND introduces $a_{_0}$ as a fundamental acceleration scale of nature below which the deviation from Newtonian dynamics becomes significant. Empirically, $a_{_0} \approx 1.2 \times {10}^{-10}$ m/s$^2$ to match galaxy rotation curves \citep{Begeman_1991}. With this value of $a_{_0}$, MOND continues to fit galaxy rotation curves very well using only their directly observed baryonic matter \citep{Li_2018, Kroupa_2018}. Moreover, it is likely that MOND can explain the Milky Way and Andromeda satellite planes as having formed out of tidal debris expelled during a past interaction between these galaxies \citep{Banik_Ryan_2018}.

MOND was originally developed for non-relativistic systems. A relativistic generalization was eventually found \citep[TeVeS,][]{Bekenstein_2004}, but this particular model has been falsified by the simultaneous detection of gravitational waves and their electromagnetic counterpart \citep{LIGO_Virgo_2017}. While these results severely constrain relativistic generalizations of MOND, they certainly do not rule out all possible ways to make MOND consistent with Lorentz invariance \citep{Sanders_2018}. Indeed, the latter work provides a simple way to do exactly that.

Although MOND was designed with galaxy data in mind, its central prediction is that departures from Newtonian dynamics arise below a particular acceleration scale rather than e.g. beyond a particular distance \citep[][figure 10]{Famaey_McGaugh_2012}. Thus, MOND can be tested in a rather small system if it has a sufficiently low mass. In particular, the MOND radius of the Sun is $r_{_M} = 7$ kAU (Equation \ref{Deep_MOND_limit}). This is much smaller than the 268 kAU distance between the Sun and its nearest star, Proxima Centauri \citep{Gaia_2018}. Therefore, interstellar missions will necessarily probe the MOND regime.

It has already been pointed out that such long-distance missions could be used to constrain modified gravity theories \citep{Christian_2017}. Those authors considered a spacecraft travelling at $0.2 \, c$, where $c$ is the speed of light in vacuum. Here, we consider an interstellar precursor mission (IPM) travelling at only $0.01 \, c$. It is not useful to launch such a mission specifically to reach our nearest stellar system as this would take 420 years. During this time, technological progress would very likely enable much faster launches, allowing future spacecraft to overtake earlier ones \citep{Heller_2017}.

Nonetheless, the extreme technical difficulties of true interstellar missions imply that we must first fly IPMs of some kind. A socio-political problem with such missions is the lack of obvious targets with distances beyond 100 AU and below the 270 kAU distance to the Sun's nearest star, Proxima Centauri \citep{Gaia_2018}. Missions reaching 550 AU could use the Sun as a gravitational telescope to magnify objects behind it \citep{Eshleman_1979}, but this idea runs into several practical difficulties for imaging exoplanets \citep{Willems_2018}.

Fortunately, the MOND radius of the Sun provides a scientifically important intermediate goal of 7 kAU, especially if efforts to directly detect cold dark matter continue to rule out ever more of the available parameter space \citep[e.g.][]{Hoof_2019}. In MOND, the trajectory of a spacecraft would start to deviate from Newtonian mechanics as it approaches a distance of $r_{_M}$. These deviations would be more significant at larger radii, though they would be limited by the background gravitational field from the rest of our Galaxy due to the external field effect \citep{Milgrom_1986, Banik_2015}. Nonetheless, significant deviations are expected once the latest extragalactic rotation curve constraints are imposed on the MOND interpolating function between the Newtonian and Milgromian regimes \citep[][section 7.1]{Banik_2018_Centauri}.

In the short run, constraints on MOND from an IPM are unlikely to be as precise as constraints obtained in other ways, in particular from extragalactic rotation curves. Testing MOND with an IPM thus serves a different purpose $-$ to test whether gravity does in fact depart from Newtonian expectations at low accelerations. If one assumes that it does, rotation curve constraints are a more promising way of constraining the limited number of MOND free parameters.\footnote{The `sharpness' of the interpolating function and the value of $a_{_0}$.} However, it is generally possible to explain rotation curve data using an appropriately tuned distribution of dark matter, even when the kinematic and photometric data come from different galaxies \citep{Blok_1998}.

This degeneracy can be broken by an IPM because its downrange distance, though large by terrestrial standards, is still very small by Galactic standards.\footnote{20 kAU $\approx 0.1$ pc while the Galaxy's virial radius is $\approx 200$ kpc \citep{Dehnen_2006}.} Because the purported acceleration due to the Galactic dark matter halo is of order $a_{_0}$, dark matter would have a negligible effect on the trajectory of an IPM compared to the effects of MOND, which has an order unity effect on the Sun's gravitational field beyond its MOND radius (Section \ref{MOND_force_field}).

We therefore consider $0.01 \, c$ IPMs travelling at various angles to the direction of the Galactic Centre, which we suppose is the direction of the external gravitational field on the Solar System. At this speed, a spacecraft would reach the MOND radius of the Sun in just over 11 years. Compared to true interstellar missions, the lower speed of an IPM gives the Sun more time to gravitationally decelerate the spacecraft, making it more sensitive to MOND effects.

An important aspect of our results is that the Sun's Milgromian gravity does not point directly towards it. This has been shown analytically in the external field dominated regime \citep{Banik_2015}. The force can be offset from the radial direction by as much as ${16^\circ}$ (see their figure 1). Consequently, an IPM would deviate from its original launch direction in a characteristic way, something that could potentially be detected by launching several missions in different directions. This would also allow us to better distinguish gravitational forces from other effects like accretion of dust and gas. For instance, the MOND gravitational field should be axisymmetric about the direction towards the Galactic Centre. This is in a rather different direction to the velocity of the Sun with respect to the interstellar medium \citep{Francis_2014}. Nonetheless, there is in principle no reason why multiple missions are necessary as one could utilize the characteristic time dependence of the MOND effects within a single mission.

After providing an order of magnitude estimate of how MOND affects an IPM (Section \ref{Analytic_estimates}), we briefly revisit the MOND governing equations (Section \ref{Governing_equations}) and explain how we solve them to yield the Milgromian gravitational field of the Sun (Section \ref{MOND_force_field}). We use this to integrate the trajectory of a spacecraft (Section \ref{Trajectory_integration}) in order to determine how much the MOND predictions differ from the Newtonian ones regarding its radial velocity (Section \ref{Radial_velocity}), downrange distance (Section \ref{Distance}) and sky position (Section \ref{Sky_position}). We then consider the feasibility of detecting the Milgromian perturbations to the spacecraft trajectory (Section \ref{MOND_detectability}) and how additional onboard experiments could be used to more directly constrain the behaviour of gravity at low accelerations (Section \ref{Additional_experiments}). We also discuss how our results would be affected by a different launch velocity (Section \ref{Different_v_launch}) and MOND formulation (Section \ref{MOND_formulations}). In Section \ref{Complementary_constraints}, we explain how an IPM would place constraints on MOND that complement those obtained using more traditional methods. Our conclusions are given in Section \ref{Conclusions}.

\section{Methods and results}
\label{Methods}

\subsection{Order of magnitude estimates}
\label{Analytic_estimates}

Newtonian and Milgromian dynamics predict gravitational accelerations which differ by $\approx a_{_0}$ at the MOND radius. Given a typical mission timescale of ${\approx 10}$ years to reach this distance, we expect the velocity of an IPM to differ by ${\approx 4}$ cm/s between the different gravity theories. Over 10 years, this translates into a position difference of 12000 km. Consequently, MOND predicts a ${\approx 0.1}$ second reduction in the two-way light travel time between the inner Solar System and the spacecraft. If 10\% of its MOND-induced position anomaly is orthogonal to the original launch direction due to the non-radial gravity in MOND (Section \ref{Sky_position}), we expect an angular deviation of ${\approx 0.2}$ mas. In the rest of this section, we refine these estimates using rigorous calculations of the Sun's gravitational field in MOND.

\subsection{Governing equations}
\label{Governing_equations}

In this contribution, our MOND predictions are based on the quasilinear formulation of MOND \citep[QUMOND,][]{QUMOND}. QUMOND yields rather similar results to the original aquadratic Lagrangian version of MOND \citep[AQUAL,][]{Bekenstein_Milgrom_1984}, as demonstrated numerically in \citet{Candlish_2016} and analytically in \citet{Banik_2015}. QUMOND is much more computer-friendly because it avoids a non-linear grid relaxation stage.

In QUMOND, the gravitational field $\bm{g}$ must be obtained from the Newtonian gravitational field $\bm{g}_{_N}$ by solving the field equation
\begin{eqnarray}
	\label{QUMOND_equation}
	\nabla \cdot \bm{g} ~&=&~ \nabla \cdot \left[\nu \overbrace{\left( \frac{g_{_N}}{a_{_0}} \right)}^y \, \bm{g}_{_N} \right] ~~\text{ , where} \\
	\nu \left( y \right) &=& \frac{1}{2} ~+~ \sqrt{\frac{1}{4} + \frac{1}{y}} \, .
\end{eqnarray}
Here, we have used the simple interpolating function $\nu \left( y \right)$ \citep{Famaey_Binney_2005} for reasons explained in section 7.1 of \citet{{Banik_2018_Centauri}}. This is numerically very similar to the function used by \citet{Lelli_2017} to fit 153 disk galaxy rotation curves.

The Newtonian gravity is rather easily obtained for a point mass (the Sun) embedded in a constant external field $\bm{g}_{ext}$. At some heliocentric position $\bm{r}$,
\begin{eqnarray}
	\bm{g}_{_N} ~=~ -\frac{GM \bm{r}}{\left| \bm{r}\right|^3} ~+~ \bm{g}_{_{N, ext}} \, .
\end{eqnarray}
The external field entering this equation is the Newtonian-equivalent quantity $\bm{g}_{_{N, ext}}$, the Newtonian gravity exerted by the rest of our Galaxy on the Solar System. As discussed in section 9.3.1 of \citet{Banik_2018_Centauri}, the Sun's location in the outskirts of the Galactic disk means that it is quite accurate to use the spherically symmetric relation between $\bm{g}_{_{N, ext}}$ and the actual external field $\bm{g}_{ext}$.
\begin{eqnarray}
	\bm{g}_{ext} ~=~ \nu \left(\frac{\left| \bm{g}_{_{N, ext}} \right|}{a_{_0}} \right) \bm{g}_{_{N, ext}} \, .
	\label{g_N_ext}
\end{eqnarray}

Kinematic observations of our Galaxy imply a particular value for $\bm{g}_{ext}$, which we get from the Local Standard of Rest parameters obtained by \citet{McMillan_2017}. We use this to get $\bm{g}_{_{N, ext}}$ by analytically inverting Equation \ref{g_N_ext}.\footnote{For more complicated interpolating functions such as that used in \citet{Lelli_2017}, analytic inversion is not possible. Equation \ref{g_N_ext} must then be inverted using a numerical root-finding algorithm.}

\subsection{The MOND force field}
\label{MOND_force_field}

After finding $\nabla \cdot \bm{g}$ using Equation \ref{QUMOND_equation}, we use direct summation to find $\bm{g}$ itself.
\begin{eqnarray}
\bm{g} \left( \bm{r} \right) ~=~ \int \nabla \cdot \bm{g} \left( \bm{r'}\right) \frac{\left( \bm{r} - \bm{r'} \right)}{4 \mathrm{\pi} |\bm{r} - \bm{r'}|^3} \,d^3\bm{r'} \, .
\label{g_direct_sum}
\end{eqnarray}
We apply an analytic correction for the finite extent of our grid, which we set up using spherical polar co-ordinates to best exploit the axisymmetric nature of the problem \citep[section 2.1,][]{Banik_2018_Centauri}.

\begin{figure}
	\centering
	\includegraphics[width = 8.5cm] {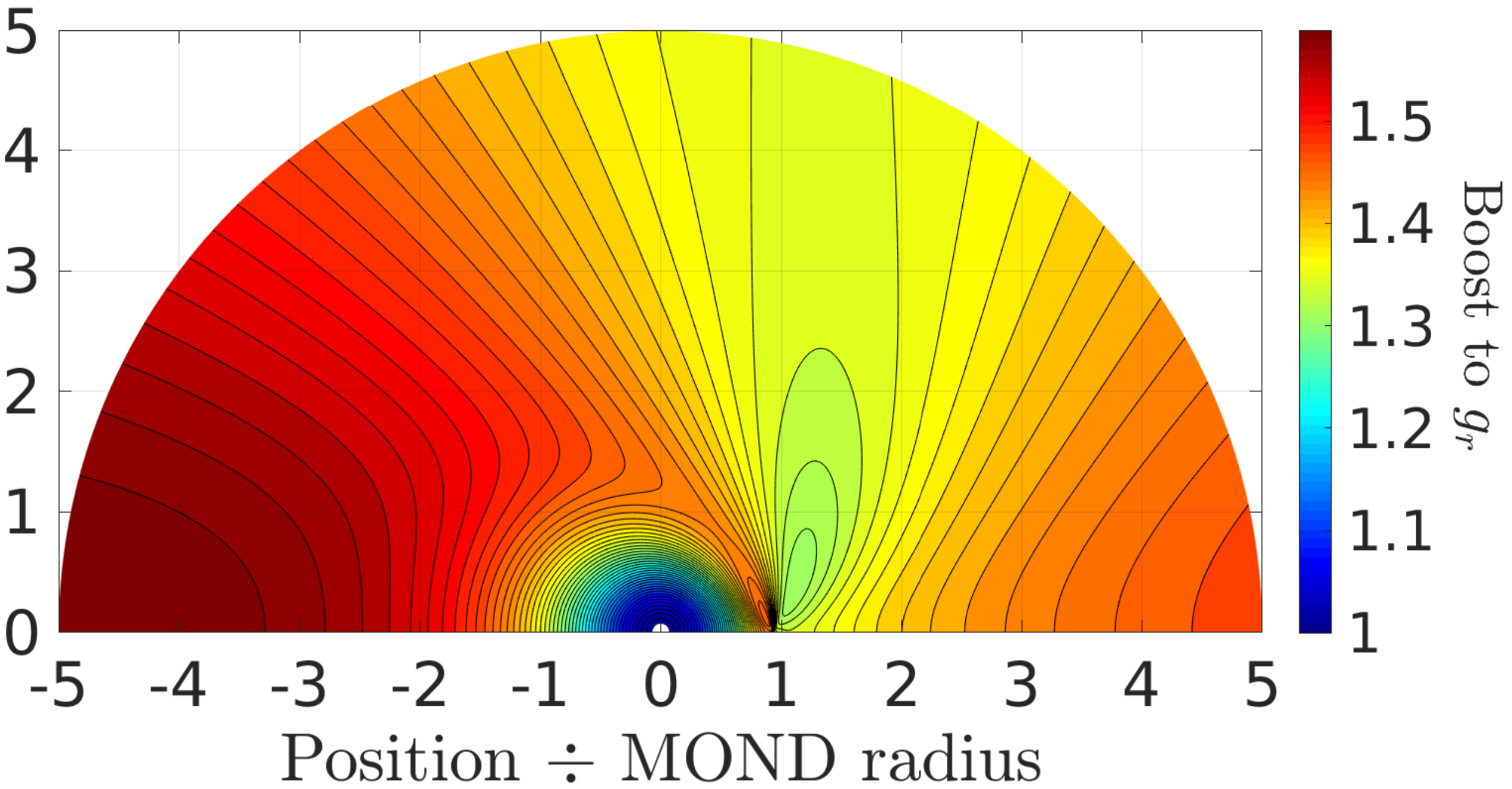}
	\caption{The factor by which MOND boosts the radial component of the Sun's gravity above the Newtonian expectation. Positions are shown in units of the MOND radius $r_{_M}$ (Equation \ref{Deep_MOND_limit}). The gravitational field is expected to be axisymmetric about the direction towards the Galactic Centre, which we represent here using the $x$-axis (the Galactic Centre is towards positive $x$). The distance orthogonal to this direction is shown on the $y$-axis. For numerical reasons, we could not obtain accurate results within $\approx 0.1 \, r_{_M}$, a region where Newtonian and Milgromian gravity should be almost identical.}
	\label{Force_ratio}
\end{figure}

In general, $\bm{g}$ is not directed towards the Sun, a fact which is easily demonstrated analytically in both QUMOND and AQUAL \citep{Banik_2015}. We use Figure \ref{Force_ratio} to show the factor by which QUMOND enhances the radial component of $\bm{g}$ at different positions. Due to axisymmetry, we only need to show our results as a function of the distance along and orthogonal to the external field direction. These variables are our $x$ and $y$ co-ordinates, respectively, with the Galactic Centre located towards positive $x$.

\begin{figure}
	\centering
	\includegraphics[width = 8.5cm] {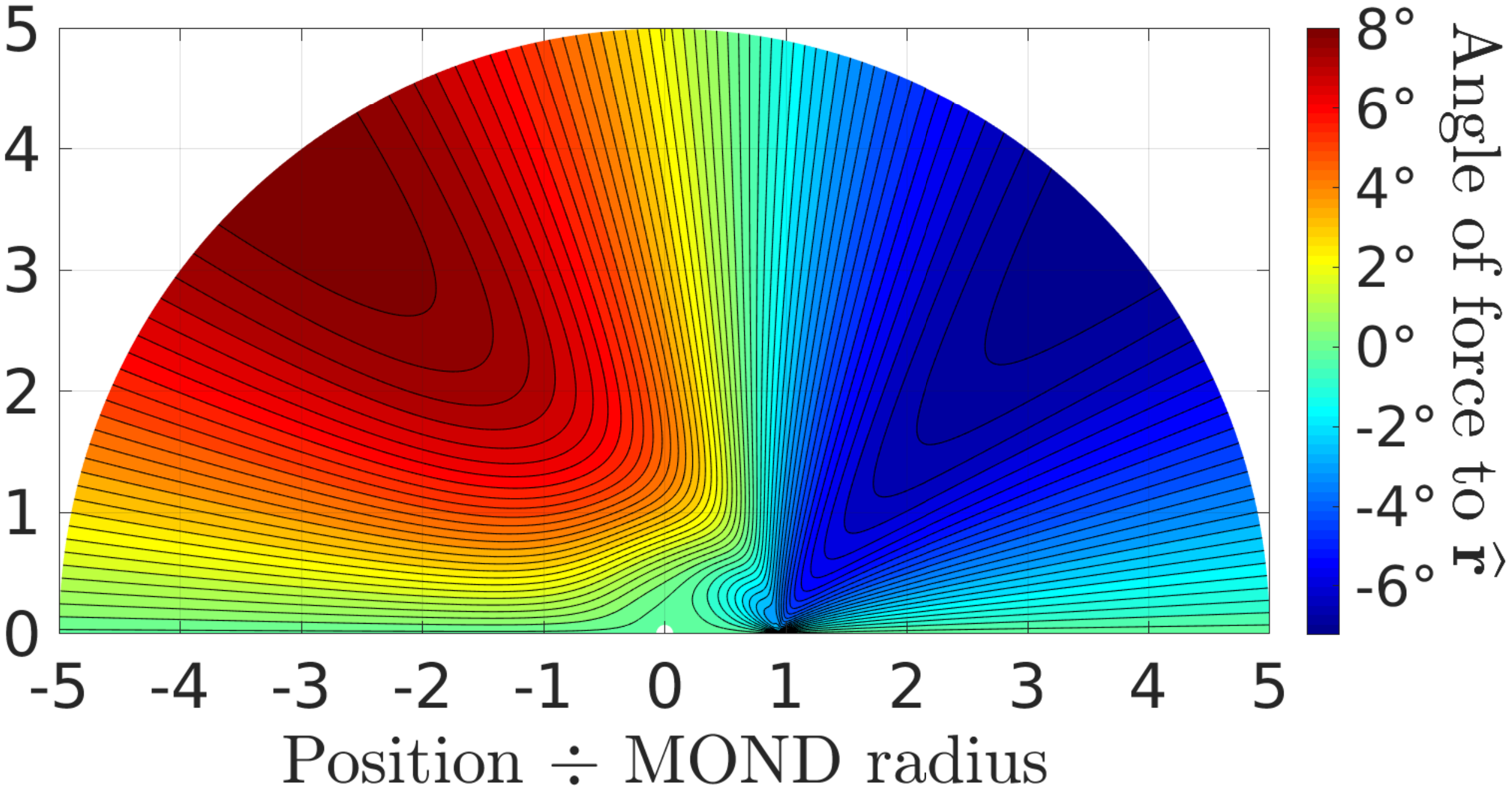}
	\caption{Similar to Figure \ref{Force_ratio}, but now showing the angle between the Sun's Milgromian gravity and the inwards radial direction. Positive angles imply the spacecraft would deviate away from the direction towards the Galactic Centre (positive $x$-axis).}
	\label{Force_angle}
\end{figure}

An important part of this contribution is the non-zero angle between the gravitational field and the inwards radial direction (Section \ref{Sky_position}). This angle is shown in Figure \ref{Force_angle} as a function of position, using the same projection system as Figure \ref{Force_ratio}. A positive angle implies a particle on a radial trajectory is deviated away from the direction towards the Galactic Centre.

\subsection{Trajectory integration}
\label{Trajectory_integration}

To avoid numerical uncertainties in the MOND gravity very close to the Sun, we only consider that portion of the spacecraft trajectory which has a downrange distance ${> 2}$ kAU, corresponding to an acceleration $< 12.3 \, a_{_0}$. Deviations from Newtonian gravity must be rather small at higher accelerations in order to satisfy Solar System constraints \citep{Hees_2016}. Moreover, the time an IPM spends within 2 kAU of the Sun constitutes only a small fraction of the time required to reach $r_{_M} = 7.03$ kAU.

At a speed of $0.01 \, c$, a spacecraft is hardly affected by the Sun's gravitational field. Therefore, we assume the spacecraft maintains a constant velocity $\bm{v}_{_{launch}}$ beyond 2 kAU but keep track of the Milgromian gravitational acceleration $\bm{g}_{_M} \left( \bm{r} \right)$ acting on the spacecraft at heliocentric position $\bm{r}$. As long as this has only a small effect on the trajectory, our perturbative approximation remains valid. This is the case here because the escape velocity of the Sun is $\approx 3 \times 10^{-6} \, c$ at a distance of 2 kAU.

\subsection{Radial velocity}
\label{Radial_velocity}

Once we have found $\bm{g}_{_M} \left( \bm{r} \right)$, we subtract the Newtonian acceleration $\bm{g}_{_N} \left( \bm{r} \right)$ and integrate the difference in accelerations over time $t$, which we measure from when the spacecraft first crosses 2 kAU at some position $\bm{r}_{_{2\,kAU}}$. This lets us determine the Milgromian perturbations to the position and velocity, which we denote $\delta \bm{r}$ and $\delta \bm{v}$, respectively.
\begin{eqnarray}
	\delta \bm{v} \left( t \right) ~&=&~ \int_0^t \left[ \bm{g}_{_M} \left( \bm{r} \right) - \bm{g}_{_N} \left( \bm{r} \right) \right] dt'\, ,\\
	\bm{r} \left( t \right)~&=&~ \bm{r}_{_{2\,kAU}} \, + \, \bm{v}_{_{launch}} t \, .
\end{eqnarray}

The line of sight component of $\delta \bm{v}$ causes an anomalous shift in the frequency of signals received from the spacecraft. Assuming its trajectory is not much affected by gravity, we get that the radial velocity $v_r$ differs from Newtonian expectations by
\begin{eqnarray}
	\delta v_r ~=~ \delta \bm{v} \cdot \widehat{\bm{v}}_{_{launch}} \, ,
\end{eqnarray}
where we use $\widehat{\bm{a}}$ to denote the unit vector parallel to $\bm{a}$ for any vector $\bm{a}$.

\begin{figure}
	\centering
	\includegraphics[width = 8.5cm] {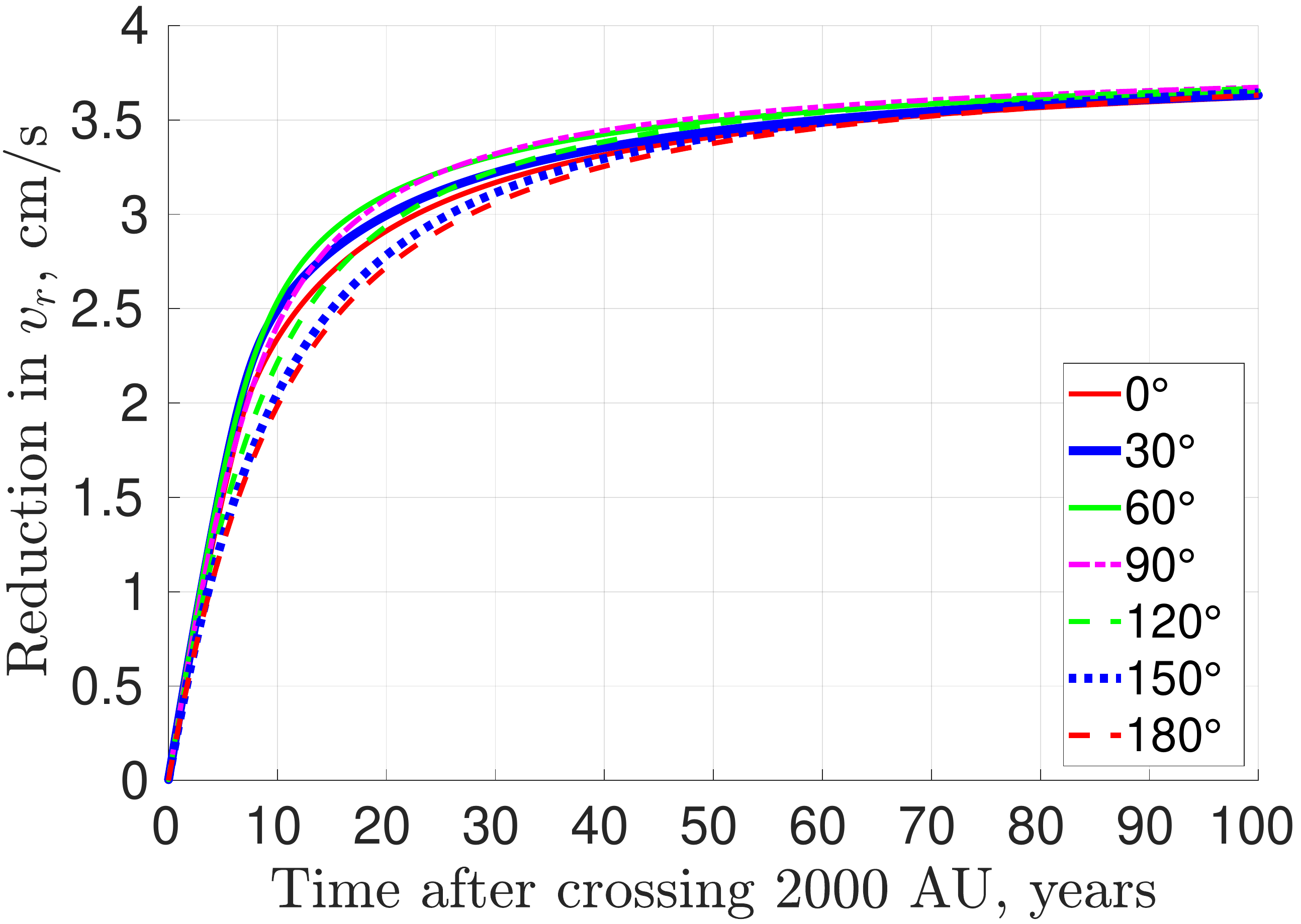}
	\caption{The difference between Newtonian and Milgromian predictions for the heliocentric radial velocity of a $0.01 \, c$ interstellar precursor mission, shown as a function of time after it crosses 2 kAU from the Sun and thus enters the regime where MOND would have non-negligible effects. The different curves correspond to different launch directions $\theta$ (Equation \ref{Angular_deviation_equation}). We use solid lines only for $\theta < \frac{\pi}{2}$ and use the same colour for $\theta$ as for $\pi - \theta$.}
	\label{Radial_velocity_time}
\end{figure}

Figure \ref{Radial_velocity_time} shows our results for $\delta v_r$. As expected, this is below 0, a consequence of the stronger gravity in MOND. The precise value is rather similar to our order of magnitude estimate in Section \ref{Analytic_estimates}.

\subsubsection{Gas drag}
\label{Gas_drag}

Assuming an interstellar medium density of ${\approx 0.2}$ proton masses per cm$^3$ \citep{Crawford_2011}, the total amount of mass accreted by a 1 m$^2$ object over 10 kAU is ${5 \times 10^{-7}}$ kg. If the spacecraft has a mass of 1 kg, the effect on its velocity would be 1.5 m/s, more than the expected effect of MOND (Figure \ref{Radial_velocity_time}). However, gas drag would be present at all distances, whereas the expected Milgromian $\delta v_r$ arises mainly around $r_{_M}$ and saturates once the spacecraft is well beyond $r_{_M}$ (Figure \ref{Radial_velocity_time}). This is because there is a finite difference in the depth of the Solar potential well between Newtonian and Milgromian dynamics \citep{Banik_2018_escape}.

Another distinction between gravity and gas drag is that the latter is less significant at lower velocity, something that could be exploited by launching multiple spacecraft at different velocities towards a similar direction. Moreover, a particle detector on board the spacecraft could be used to measure the accretion rate, aiding studies of the interstellar medium. The detector would need to withstand high impact velocities of ${\approx 3000}$ km/s.

\subsection{Downrange distance}
\label{Distance}

The change in velocity would build up over many years into a potentially detectable position perturbation
\begin{eqnarray}
	\delta \bm{r} \left( t \right) ~&=&~ \int_0^t \delta \bm{v} \left( t' \right) dt' \, .
\end{eqnarray}
This could be measured by determining the downrange distance $r \equiv \left| \bm{r} \right|$ of the spacecraft. Neglecting the second-order effect of the angular deviation (Section \ref{Sky_position}) and assuming the spacecraft is tracked from a location much closer to the Sun than $r$, we get that
\begin{eqnarray}
	\delta r ~=~ \delta \bm{r} \cdot \widehat{\bm{v}}_{_{launch}} \, .
\end{eqnarray}
This changes the two-way light travel time from the inner Solar System by
\begin{eqnarray}
	\delta t_{2 \, way} ~=~ \frac{2 \, \delta r}{c} \, .
	\label{Delta_t_2_way}
\end{eqnarray}

\begin{figure}
	\centering
	\includegraphics[width = 8.5cm] {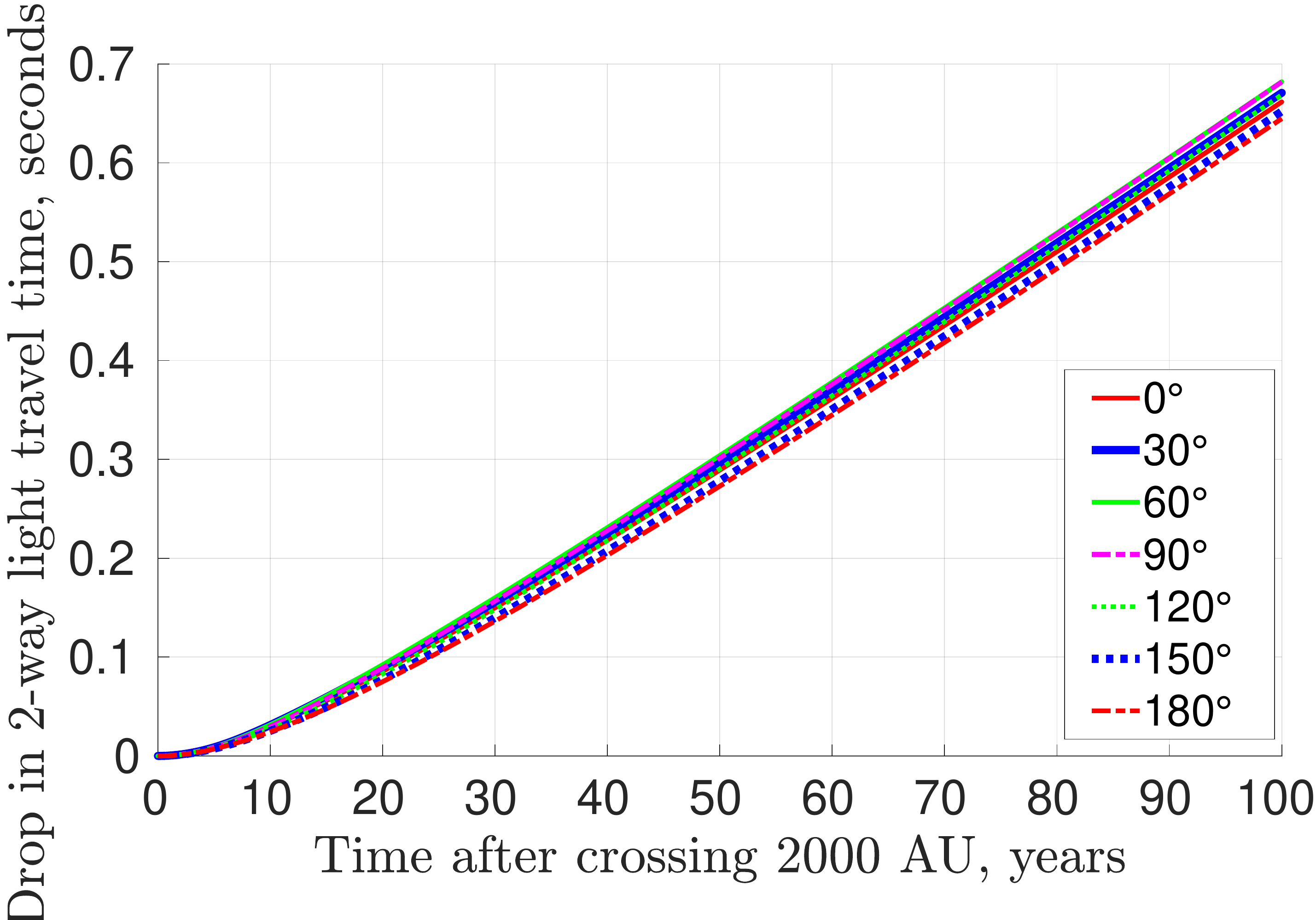}
	\caption{Similar to Figure \ref{Radial_velocity_time}, but now showing how much MOND would reduce the two-way light travel time between the inner Solar System and the spacecraft.}
	\label{Time_delay_time}
\end{figure}

Figure \ref{Time_delay_time} shows that $\delta t_{2 \, way}$ depends very little on $\widehat{\bm{v}}_{_{launch}}$ and is ${\approx 0.1}$ seconds after 20 years. Once the spacecraft is well beyond the MOND radius, $\delta t_{2 \, way}$ grows approximately linearly with time because $\delta v_r$ saturates (Figure \ref{Radial_velocity_time}).

\subsection{Sky position}
\label{Sky_position}

Another MOND prediction is that the spacecraft trajectory would deviate from its original direction $\widehat{\bm{v}}_{_{launch}}$ because $\bm{g}_{_M} \left( \bm{r} \right)$ is not anti-parallel to $\bm{r}$ (Figure \ref{Force_angle}). This is due to the Sun's Milgromian potential varying with angle at fixed heliocentric distance. In the asymptotic regime where Solar gravity is much weaker than that from the rest of the Galaxy, the Solar potential is deepest along the Galactic Centre-anticentre line \citep{Banik_2015}.

Observations from the inner Solar System are not directly sensitive to the velocity vector of such a distant spacecraft, but over time this affects its position on the sky. Thus, we define its angular deviation $\delta \theta$ as the angle between $\widehat{\bm{v}}_{_{launch}}$ and the instantaneous direction $\widehat{\bm{r}}$ towards the spacecraft as viewed from the Sun. Because $\bm{g}_{_M}$ is axisymmetric with respect to the direction $\widehat{\bm{GC}}$ towards the Galactic Centre, the angular deviation is entirely within the plane containing $\widehat{\bm{GC}}$ and $\widehat{\bm{r}}$. If MOND increases the angle $\theta$ between them, we take $\delta \theta$ to be positive.
\begin{eqnarray}
	\delta \theta ~=~ \frac{\delta \bm{r} \cdot \left[\overbrace{\left( \widehat{\bm{GC}} \cdot \widehat{\bm{v}}_{_{launch}} \right)}^{\cos \theta} \widehat{\bm{v}}_{_{launch}} \, - \, \widehat{\bm{GC}}\right]}{r \sin \theta} \, .
	\label{Angular_deviation_equation}
\end{eqnarray}

\begin{figure}
	\centering
	\includegraphics[width = 8.5cm] {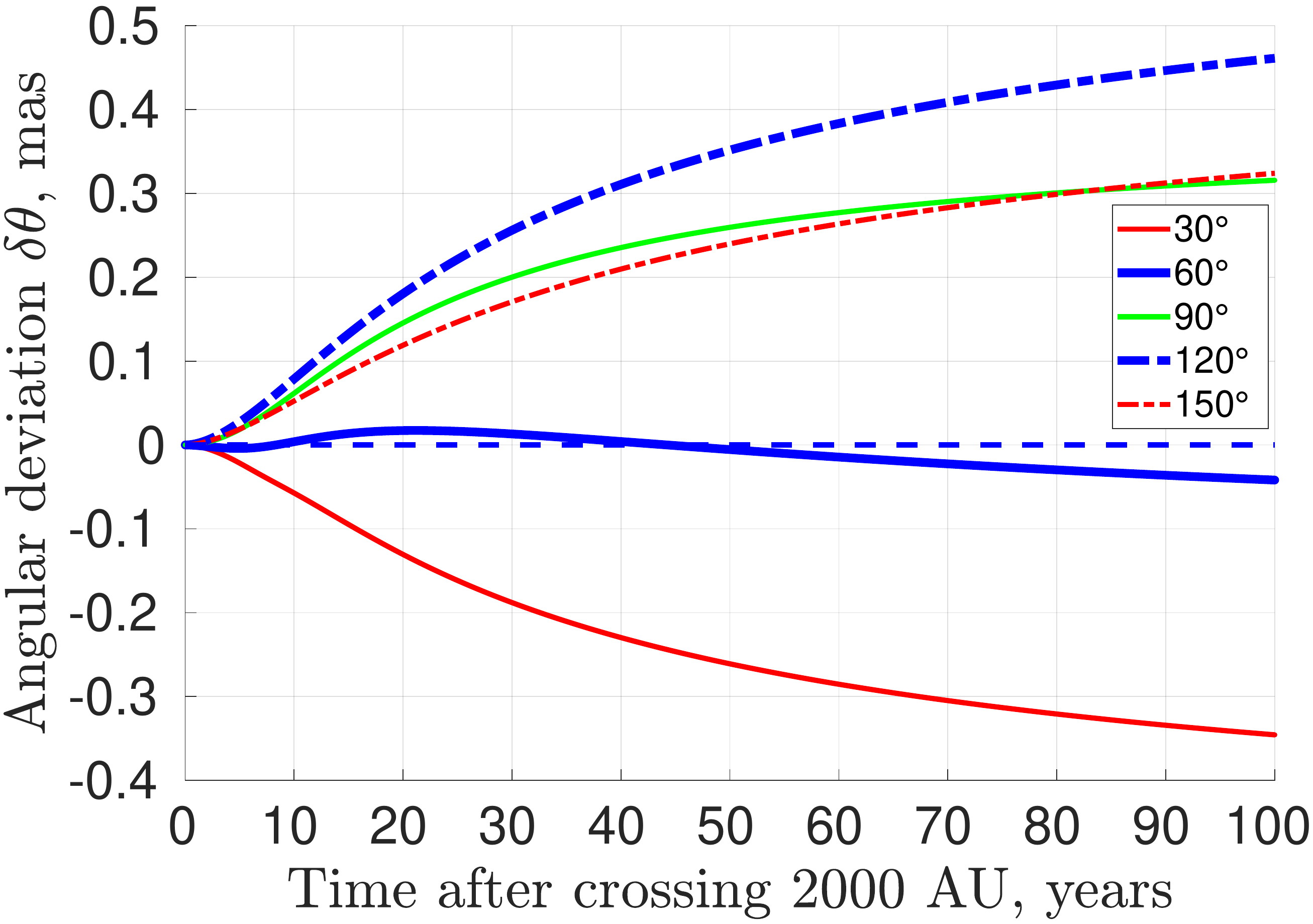}
	\caption{Similar to Figure \ref{Time_delay_time}, but now showing the MOND prediction for the change in direction towards the spacecraft as a consequence of the Sun's gravity not being directly towards it (Equation \ref{Angular_deviation_equation}). The horizontal dashed blue line shows the Newtonian prediction of zero, which also applies in MOND for launch directions of ${\theta = 0}$ and ${\pi}$.}
	\label{Angular_deviation_time}
\end{figure}

We use Figure \ref{Angular_deviation_time} to show our results for $\delta \theta$ as a function of time for various launch directions. In terms of maximizing ${\delta \theta}$, the best choices are $\theta \approx 30^\circ$ or $120^\circ$. Each option can be implemented for any azimuthal angle with respect to $\widehat{\bm{GC}}$, allowing mission planners to choose whichever is best in light of other considerations. For example, the mission could be designed to have suitable reference stars in the same field of view as the Sun (Section \ref{Onboard_astrometry}).

Some asymmetry is evident between $\theta$ and $\pi - \theta$, highlighting the need for careful numerical calculations. This asymmetry was a key argument in \citet{Thomas_2018}. Once the external field from the Galaxy dominates over Solar gravity, $\bm{g}_{_M} \left( \bm{r} \right)$ becomes symmetric with respect to $\theta \to \pi - \theta$ \citep[][equation 36]{Banik_2015}. Theoretically, this transition occurs close to $r_{_M}$ because the Galactic gravity ${\approx a_{_0}}$. Our results indicate a rather gradual transition to a symmetric gravitational field, something that is also evident in Figures \ref{Force_ratio} and \ref{Force_angle}.

\section{Discussion}

\subsection{Detectability of the Milgromian deviations}
\label{MOND_detectability}

\subsubsection{Radial velocity and downrange distance}
\label{Delta_RV_detectability}

The expected Milgromian boost to the Sun's radial gravity would reduce the downrange distance to an IPM by ${\approx 15000}$ km 20 years after launch (Figure \ref{Time_delay_time}). Based on accurate tracking of the Cassini orbiter around Saturn \citep{Matson_1992}, its position was known to an accuracy of 32 m \citep{Viswanathan_2017}. The interpretation of light travel times from an IPM should be vastly simpler because Cassini was affected by the gravity of Saturn and its many moons in addition to the Solar gravity. Therefore, the expected MOND-induced distance reduction should be readily detectable.

\subsubsection{Sky position}
\label{Delta_theta_detectability}

An important aspect of this work is that in MOND, an IPM is predicted to deviate from its original launch direction (Figure \ref{Angular_deviation_time}). The expected ${\approx 0.1}$ mas signal corresponds to $\delta \theta = 5 \times 10^{-10}$ radians. This could be detected by noting the times at which a signal from the spacecraft arrives at different receiving stations. For two stations separated by some distance $L$, the difference in downrange distances to the spacecraft is affected by $\approx L \, \delta \theta$.\footnote{The exact geometry introduces an additional factor ${\leq 1}$.}

By the time that IPMs are actually flown, it is quite conceivable that a station on the Earth will be combined with one on the Moon \citep{Kurdubov_2019}. Given that its distance is known to mm accuracy using lunar laser ranging \citep{Battat_2009}, the Moon could be a good location for a receiving station immune to atmospheric effects. Because the Moon is $L \approx 4 \times 10^8$ m away, the MOND-induced astrometric deviation would have a ${\approx 20}$ cm effect on the difference between the Earth-spacecraft and Moon-spacecraft distances. This should be readily measurable. In fact, it is even possible that different receiving stations on Earth would be sufficient.

Another possibility is to combine measurements taken at different times, exploiting Earth's motion around the Sun. This would require precise knowledge of its orbit and rotation, at least on short timescales. For instance, Earth's orbit moves it by the distance to the Moon in just under 4 hours.

The downrange distance to an IPM can be known to within 20 cm only if this is much longer than the wavelength at which the spacecraft transmits. A wavelength of 20 cm corresponds to a frequency of 1.5 GHz, similar to the frequencies typically used for communication with spacecraft. However, it is unlikely that radio waves will be used for communication with an IPM due to its much larger downrange distance. We consider it likely that a visible or near-infrared laser-based system will be used instead, as demonstrated recently by the Lunar Environment and Dust Explorer mission \citep{Boroson_2014}. In this case, the transmission wavelength would be of order $1 \, \mu$m, much smaller than the 20 cm precision required to test MOND.

\subsection{Additional experiments outside the Sun's MOND bubble}
\label{Additional_experiments}

So far, we have only considered information which can be gained from tracking the spacecraft. Depending on technological constraints, it is possible to envisage additional onboard experiments. For instance, a Terrell interferometer \citep{Terrell_1959, Penrose_1959} could be used to measure its acceleration more precisely \citep{Christian_2017}. Any non-gravitational forces acting on the spacecraft could be measured by an accelerometer \citep{Lenoir_2011}, possibly avoiding the issues that arose with the Pioneer anomaly \citep{Turyshev_2012}. Gas drag could be estimated using a detector, which would also give insights into the interstellar medium.

\subsubsection{Accurate astrometry of the Sun}
\label{Onboard_astrometry}

The Milgromian angular deviation of the spacecraft trajectory (Section \ref{Sky_position}) is comparable to the astrometric accuracy of the Gaia mission \citep{Gaia_2018}. Therefore, this deviation could be measured by the spacecraft itself if it is capable of attaining Gaia-like astrometry of the Sun. This may prove to be easier than ground-based tracking of the spacecraft (Section \ref{Delta_theta_detectability}), simply because its signal will be so much fainter than the Sun. From the vantage point of the spacecraft, the Sun will be in almost the same sky position for many years. This would allow many different images of the Sun to be combined, thereby yielding a rather precise estimate of its sky position relative to background stars. To limit the amount of data transferred to Earth, much of the analysis could be done onboard. For example, the spacecraft might only transmit data for small regions centred on target stars, including the Sun. The launch direction could even be chosen so as to have suitable reference stars in the field of view.

Although a Gaia-like telescope onboard the spacecraft would make it much heavier, the benefits to science could well be worthwhile. In particular, the telescope could be used to obtain accurate astrometry of other stars besides the Sun. Due to its significant downrange distance, the positions of nearby stars would be noticeably different than if viewed from Earth. This could allow for rather accurate distance measurements to stars in our Galaxy and perhaps also for more distant objects \citep{Jaffe_1979, Etchegaray_1987}. Given the already impressive results obtained by Gaia \citep{Perryman_2001} with a 2 AU baseline, the results with a 10 kAU baseline could indeed be remarkably accurate once relativistic aberration is properly accounted for. The spacecraft velocity can be inferred from onboard astrometry, but it might also be better to use arrival times of signals sent from the spacecraft (Section \ref{Delta_theta_detectability}) and perhaps also their frequency.

\subsubsection{Onboard gravitational experiments}
\label{Cavensish_experiments}

If a slightly heavier spacecraft can be launched, it might be possible to perform Cavendish-style gravitational experiments beyond the Sun's MOND radius. As discussed in \citet[][figure 1]{Banik_2018_Centauri}, the force between two point masses would exceed the Newtonian expectation by ${\approx 30-60\%}$ depending on their heliocentric distance, orientation with respect to the Galactic external field and, to a smaller extent, their mass ratio (see their figure 8). In general, the force would not be parallel to the separation between the masses, causing the system as a whole to spontaneously undergo torsional oscillations. These unusual effects should be easily discernible given that laboratory experiments routinely obtain acceleration measurements precise to ${\ll a_{_0}}$. For instance, the detection of 35 Hertz gravitational waves with a strain amplitude of ${10^{-21}}$ using 4 km long arms \citep{LIGO_2016} implies an acceleration measurement accurate to $\la 10^{-14}$ m/s$^2$. This is much smaller than the ${\approx 0.6 \, a_{_0}}$ gravitational acceleration on a test mass due to a 1 kg object 1 m away. At present, such experiments do not place meaningful constraints on MOND due to the external gravitational field of the Earth. However, an active gravitational experiment beyond the Sun's MOND bubble could decisively test if the anomalous rotation curves of galaxies arise from the behaviour of gravity at low accelerations.

\subsection{Effect of a different launch velocity}
\label{Different_v_launch}

In the perturbative approximation used here, we can easily scale our results to different $v_{_{launch}} \equiv \left| \bm{v}_{_{launch}} \right|$ as long as gravity has only a small effect on the trajectory. If $v_{_{launch}}$ is scaled by some factor $\alpha$ from our nominal assumption of $0.01 \, c$, the Milgromian perturbations to the trajectory at time $t'$ are related to those of the original trajectory at time $t$ according to
\begin{eqnarray}
	\delta \bm{v}' \left( t' \right) ~&=&~ \frac{\delta \bm{v} \left( \alpha t \right)}{\alpha} \, , \\
	\delta \bm{r}' \left( t' \right) ~&=&~ \frac{\delta \bm{r} \left( \alpha t \right)}{\alpha^2} \, , \\
	\delta \theta' \left( t' \right) ~&=&~ \frac{\delta \theta \left( \alpha t \right)}{\alpha} \, .
	\label{v_launch_scaling}
\end{eqnarray}
Primed quantities indicate results for the trajectory with a velocity of $\alpha \, \bm{v}_{_{launch}}$ at a heliocentric distance of ${r = 2}$ kAU.

While the spacecraft is within $r_{_M}$, our results show that $v_{_{launch}}$ does not much affect $\delta v_r$ at fixed $t$ (Figure \ref{Radial_velocity_time}). As the mission progresses to larger radii, $\delta v_r$ eventually reaches a fixed value. In this regime, $\delta v_r \propto 1/v_{_{launch}}$, causing $\delta r$ to eventually be much smaller at higher launch velocities. This underlines how the slower speeds of IPMs actually benefit the test of gravity proposed here. In particular, reducing $v_{_{launch}}$ minimises gas drag while giving gravity more time to act on the spacecraft.

A unique aspect of MOND is that it predicts a change in the sky position of the spacecraft (Section \ref{Sky_position}). Early in a mission, the Milgromian $\delta \theta$ grows roughly linearly with time. Thus, the launch velocity has little effect on $\delta \theta$ at fixed $t$ (Equation \ref{v_launch_scaling}). However, $\delta \theta$ eventually grows sub-linearly with time such that it is larger at fixed $t$ for a slower launch.

In all cases, there are at most modest benefits to a faster launch during the early phases of the mission. But at later stages, this actually makes the signal smaller. Thus, a launch velocity slightly below $0.01 \, c$ may be best for the purpose of testing MOND. Given that such a mission could get overtaken by a successor before it reaches $r_{_M}$, it is probably best to launch once it becomes possible to reach this goal in ${\approx 20}$ years \citep{Heller_2017}. This corresponds to a speed of $\approx 0.006 \, c = 1800$ km/s. Although this is not currently feasible, it is nonetheless interesting to consider the science that might be enabled by such a mission.

\subsection{Different MOND formulations}
\label{MOND_formulations}

Our results are based on QUMOND, a computer-friendly version of MOND which gives rather similar results to the more traditional aquadratic Lagrangian formulation \citep[AQUAL,][]{Bekenstein_Milgrom_1984}. AQUAL is less computer-friendly, so it is beyond the scope of this work to consider AQUAL in detail. However, some insights can be gained from previous analytic calculations in the external field-dominated regime, roughly corresponding to ${r \gg r_{_M}}$ for any point-like object in the Solar neighbourhood \citep[][figure 1]{Banik_2015}. Their results suggest that both formulations yield rather similar perturbations to the light travel time and radial velocity, but the angular deviation would be ${\approx 20\%}$ larger and peak for slightly higher $\left| \cos \theta \right|$ in the AQUAL formulation. The basic result is rather similar in both versions of MOND, which after all give the same forces in spherical symmetry.

\subsection{Complementary constraints on MOND}
\label{Complementary_constraints}

Because the Galactic orbital acceleration of the Sun is close to $a_{_0}$, our results are sensitive to the MOND interpolating function. Fortunately, this is rather well constrained empirically \citep{Lelli_2017}. The `simple' form used here fits their observations rather well and is preferred over other forms by a variety of observations \citep[][section 7.1]{Banik_2018_Centauri}. In future, extragalactic constraints on MOND should tighten further, leaving little room to change its interpolating function to fit tracking data from an IPM.

It is worth noting that the MOND-induced reduction in an IPM's radial velocity is mostly sensitive to the interpolating function itself. Thus, the perturbations to the radial velocity and downrange distance are almost directly related to extragalactic constraints. The angular deviation of the trajectory depends on the derivative of the interpolating function, which governs how much the Solar gravitational potential varies with angle at fixed heliocentric distance \citep[equations 20 and 39 in][]{Banik_2015}. Consequently, this is less uniquely determined from the basic tenets of MOND combined with rotation curve data. This is why the astrometric deviation is expected to differ somewhat between AQUAL and QUMOND, even when the same interpolating function is used.\footnote{For a definition of what it means to use the `same' interpolating function in AQUAL and QUMOND, we refer the reader to section 7.2 of \citet{Banik_2018_Centauri}.} An IPM thus has the potential to distinguish between QUMOND and AQUAL, even though they yield rather similar results in most circumstances \citep{Candlish_2016, Banik_2015}.

\newpage
\section{Conclusions}
\label{Conclusions}

An IPM travelling at $0.01 \, c$ would reach the MOND radius of the Sun in 11.1 years, after which time the spacecraft would experience a Solar gravitational field weaker than the $a_{_0}$ threshold of MOND. Consequently, the MOND trajectory of such a spacecraft would deviate from Newtonian expectations. After 20 years, its radial velocity would be ${\approx 3}$ cm/s lower than expected in Newtonian mechanics (Figure \ref{Radial_velocity_time}). Due to the long time span involved, the two-way light travel time would then be ${\approx 0.1}$ seconds less than expected (Figure \ref{Time_delay_time}). These values would depend on mission elapsed time in a characteristic way, potentially allowing a distinction between gravitational and non-gravitational forces like gas drag (Section \ref{Gas_drag}).

Another novel MOND prediction is that the Solar gravity on the spacecraft would not be directly towards the Sun. As a result, its sky position would appear to deviate by $\approx 0.3$ mas during the first 20 years of the mission (Figure \ref{Angular_deviation_time}). If the spacecraft is spin-stabilised with axis pointing towards the Sun, then gas drag would be unlikely to produce such an effect. Moreover, the deviation would depend on the precise direction in which the spacecraft is launched, potentially allowing for the detection of a highly distinctive MOND signature if multiple spacecraft were launched in different directions.

Because $0.01 \, c$ is much larger than the escape velocity from the Solar System, gravity has very little effect on a spacecraft travelling at such a high speed. Consequently, the test of gravity proposed here actually becomes more difficult with a higher launch velocity (Section \ref{Different_v_launch}). An exception arises if the spacecraft carries an onboard Cavendish-style experiment to directly test gravity at low accelerations (Section \ref{Additional_experiments}). Such an experiment is predicted to yield rather unusual results in a MOND context, with significant deviations expected from Newtonian dynamics. The results could be obtained earlier if the launch velocity is larger. Given past rates of technological progress, it is probably best to launch such a mission once it becomes possible to attain a velocity of ${\approx 0.006 \, c}$ \citep{Heller_2017}.

Before humanity can reach other stars, it is necessary to fly IPMs travelling at only a few percent of light speed. These missions are directly sensitive to the behaviour of gravity in the low acceleration regime typical of galactic outskirts, where large dynamical discrepancies have puzzled astronomers for nearly a century. Directly probing these regimes could provide a short-term incentive to launch such missions, in addition to the longer-term goal of reaching other stars. While this age-old dream is perhaps closer to reality than ever before, travelling even a few percent of the required distance would be sufficient to answer fundamental questions about how our Universe works.

\section*{Acknowledgements}

IB is supported by an Alexander von Humboldt postdoctoral fellowship. He is grateful for comments by the referee which greatly improved this manuscript. The algorithms were set up using \textsc{matlab}$^\text{\textregistered}$.

\bibliographystyle{mnras}
\bibliography{SPT_bbl}
\bsp
\label{lastpage}
\end{document}